\documentclass[aps,pre,onecolumn,showpacs,showkeys,a4paper]{revtex4}
\input epsf  
\usepackage{graphicx} 
\usepackage{amsmath}
\usepackage{amssymb}
\usepackage{amscd}

\begin{document} 

 \title{Current Fluctuations in the exclusion process and Bethe Ansatz}
\author{Sylvain Prolhac and Kirone Mallick}
  \affiliation{Institut  de Physique Th\'eorique, C. E. A.  Saclay,
 91191 Gif-sur-Yvette Cedex, France}
 \author{}
  \affiliation{}
  \email{}
  \date{February 4, 2008}

\begin{abstract} 
  We use  the Bethe  Ansatz to derive  analytical expressions  for the
 current  statistics in  the  asymmetric exclusion  process with  both
 forward and  backward jumps. The  Bethe equations are  highly coupled
 and  this fact  has impeded  their use  to derive  exact  results for
 finite  systems.    We  overcome  this  technical   difficulty  by  a
 reformulation of  the Bethe equations into a  one variable polynomial
 problem,  akin  to the  functional  Bethe  Ansatz.  The  perturbative
 solution of this equation leads  to the cumulants of the current.  We
 calculate here the first two orders and derive exact formulae for the
 mean value of the current and its fluctuations.
   \end{abstract}
\maketitle 

 \section{Introduction}

     The asymmetric exclusion process  (ASEP) plays   a seminal role 
  in non-equilibrium physics of low dimensional systems
  \cite{MartinRev1}.  In its simplest version, the ASEP describes
 a system of particles, randomly  hopping on a lattice with hard-core
 exclusion interaction so that  a lattice site can be
   occupied by only one  particle at a given time.  
 Due to its minimal character, this model
 appears as a building block in many  seemingly  unrelated
  fields \cite{spohnbook}.  By virtue of   different  mappings,
 the ASEP can be interpreted as a 
  model for  RNA transcription \cite{mcdonald},
  hopping conductivity, polymers in random media,
 surface growth \cite{halpinhealy},   traffic flow,
  molecular motors \cite{lipowski} 
  etc...  In the one-dimensional case,
   many  exact results have been derived for the ASEP
   (for a review, see e.g., \cite{derridareview,schutz2}). As  a result,
 the relations  between the 
 intrinsic stochasticity  of the dynamics, the external drive and the
 particles interactions are better understood.  The fact that ASEP
 in one-dimension  is  an exactly  solvable model should not be considered 
   as just an elegant   mathematical anomaly at odds with
   physical relevance.  Indeed, many 
   of the exact results obtained for ASEP have shed
  light on the behaviour of general driven diffusive systems
  by  providing us with  effective
  phenomenological descriptions that can be applied 
  to more realistic models \cite{schnitzia}. Examples of such
 descriptions that stem  from   mathematical results are:
  shock fronts  to model
   boundary induced phase transitions,  
  the interpretation of shocks as real space condensation
   (related to  zero range processes)
    \cite{MartinBraz,mustansir},  and 
  the additivity  principle~\cite{bodineau}. Besides, the ASEP
 is  a good toy-model to test the validity of general 
  claims about  non-equilibrium systems:  for example, the
  Gallavotti-Cohen  fluctuation theorem is satisfied by  the ASEP
  (and by  more general Markovian systems)
  as  can be shown  by  elementary methods \cite{lebowspohn}
  whereas the  proof for deterministic
  dynamical systems requires some  restrictive
  hypothesis and  is far  more technical.

  Exact solutions for the exclusion process have been obtained
 by using several different approaches, and in particular the Matrix Product
 representation and the  Bethe Ansatz. 
  The Matrix Product  representation
  was first introduced  in   \cite{DEHP} to study 
  the stationary state and the phase diagram 
  of the ASEP with open boundaries. The main idea
   consists in representing the stationary state  as
 a trace over a suitable, usually quadratic,  algebra;
 this technique  has been generalized
 to many different models, including systems with shock
 profiles and with different classes 
 of particles \cite{DJLS, speer,km,MMR}. An exhaustive  and pedagogical 
   review on the Matrix  method can be found in
 \cite{MartinRev2}. 
   The Bethe Ansatz was first used  to calculate the spectral gap 
  of the ASEP and the associated   dynamical exponent 
 \cite{dhar,gwaspohn,kim1,kim2,ogkm1}. Indeed,  
 the Markov   matrix of the ASEP
  that   encodes its  stochastic  evolution  can 
   be mapped exactly onto  a  non-hermitian  spin chain 
 hamiltonian which is  integrable.   
  The Bethe Ansatz  also allows to 
  study  spectral degeneracies \cite{ogkm2}, and to investigate
  variants of the  ASEP  and 
  more  general particle hopping processes 
   \cite{schutz,evans,priezzhev,povolotsky}
 (for a review see e.g.,  \cite{ogkmrev}).

  A   particularly important  physical quantity 
   in the ASEP is the statistics of the current in the
 stationary regime. This current becomes  a local   height variable
 when the ASEP is translated into  
  a Random Solid On Solid model that describes the growth
 of a random interface.  Indeed, in  this mapping, a forward
 random  jump  of a   particle through a bond corresponds  to 
 a random deposition event of a unit 'brick' on the interface;
 a backward jump  corresponds to the evaporation of a brick. 
  The  time integrated   current through a bond of the  
  ASEP  is therefore equivalent  to the total height
   of the interface at a  given point.
      In the continuous limit, the motion of this  interface
  is described by the Kardar-Parisi-Zhang (KPZ)  equation 
  (see e.g., \cite{halpinhealy}). The exclusion process in one dimension
 is thus   a discretized version of the KPZ equation and 
   exact results about the ASEP    have  therefore 
  interesting interpretations in terms of surface growth.   

   For the exclusion process on a periodic ring, the mean value
  of the stationary current through a bond can be easily  derived
   from  elementary combinatorics; in the limit of a  large system 
  the mean current is given by the density of vacancies multiplied
  by the asymmetry rate.  However, the higher moments  of the current 
  in the stationary state are much more difficult to calculate.
 In fact,   the full  statistics of the current was determined
 only for the particular case of 
  the {\it totally} asymmetric exclusion process  (TASEP),
  where  the particles are allowed to jump only in one direction. 
  For any system size, 
 an analytical expression for the cumulant generating function
 was obtained,  leading to an  exact formula for the large deviation
 function \cite{DLebowitz,appert}.  
 This result was derived  using   the Bethe equations which,
 for the TASEP, can be solved explicitely thanks to a  decoupling
 property that  reduces  them to a one variable polynomial equation
  plus  a self-consistency condition \cite{gwaspohn,evans,ogkmrev}.

 In the general  case, when jumps
 on both directions are allowed,  the Bethe equations do not decouple
 and it has not been possible to use them to derive exact results
 for finite systems. 
 An exact formula for the  fluctuation of the current ({\it i.e.}, the 
 second moment of the current)  in the
 long time limit  could however be  derived  using an extension of 
 the matrix method \cite{DEMuk,DMal}. But higher moments appeared  
 to be out of reach. 
 
 The aim of the present   work is to derive  analytical results for the
 current statistics in  ASEP  with forward
 and backward jumps (sometimes called the  partially asymmetric 
 exclusion process)  from  the Bethe Ansatz. 
 We overcome the technical difficulty that hindered the solution 
 of the Bethe equations  in the general case by 
  reducing them  to an effective one variable 
  problem thanks to  a suitable reformulation, akin to the so-called
 functional Bethe Ansatz.  This one variable
  equation  can be interpreted as a purely algebraic question 
 involving a  divisibility condition  
 between two  polynomials. 
 In this work, we use this formalism
 to  derive  the expressions
 of the mean value of the current and  its variance.  
  Our technique can be used    
 to calculate  the current cumulant  to any desired  order.

 The outline of this work is as follows. In section~\ref{sec:eigenval}, 
 we explain that the cumulant generating function can be expressed 
 as the maximal eigenvalue of a suitable deformation of the Markov
 matrix where the deformation parameter represents 
 the   fugacity of the  jumps.
  In section~\ref{sec:Bethe}, we give the Bethe equations that 
 allow to diagonalize this matrix. The reformulation
 of the  Bethe equations as a  problem in polynomial divisibility
 is done in section~\ref{sec:FBA}. In section~\ref{sec:pert},
 we solve perturbatively this purely  algebraic problem   to the
 second order with respect to the  jump fugacity. This allows us to derive 
 the exact formulae for the mean value  and the variance of the current
 in section~\ref{sec:formules}. The last section is devoted
 to concluding remarks. Some technical derivations are given in the
 appendices.

 \section{Current statistics as an eigenvalue problem}
 \label{sec:eigenval}

\subsection{The asymmetric exclusion process}

   The exclusion process   on a periodic
    one dimensional lattice with $L$ sites (sites $i$ and $L + i$ are
   identical) is a stochastic interacting particle model in which 
      each   lattice site is   occupied by at most    one particle
 at a given time ({\it exclusion rule}).
  The  system evolves  
   with  time according to a  stochastic  dynamics:
 a particle on a site $i$ at time $t$ jumps, in
the interval between  $t$ and $t+dt$, with probability $p\ dt$ to the
neighbouring site $i+1$ if this site is empty 
 and with  probability $q\ dt$ to the
 site $i-1$ if this site is empty. The jump  rates $p$ and $q$
 are normalized such that  $ p + q =1$.  The special case
 where the jumps are totally biased in one direction ($p =1$ or  $q =1$)
 is  called  the totally asymmetric exclusion process (TASEP).
  For $p=q=1/2$,  the   exclusion process is symmetric (SEP).
 If the number of particles in the ring is $N$, the total number of
 configurations is given by the binomial coefficient
 $ \left(\begin{array}{c} L\\ N \end{array} \right).$

  We call $P_t(\mathcal{C})$
 the probability  that the system is in
  the configuration $\mathcal{C}$ at time $t$.
  As the exclusion process is a continuous-time Markov process, the time
 evolution of $P_t(\mathcal{C})$ is determined by the master equation 
\begin{equation}
    \frac{d}{dt} P_t(\mathcal{C})  = \sum_{\mathcal{C}'}
      M(\mathcal{C},\mathcal{C}') P_t(\mathcal{C}') 
   =    \sum_{\mathcal{C}'}   
 \Big(  M_0(\mathcal{C},\mathcal{C}') +   M_1(\mathcal{C},\mathcal{C}')
+   M_{-1} (\mathcal{C},\mathcal{C}')\Big)
          P_t(\mathcal{C}')   \, . 
 \label{eq:Markov}
\end{equation}
 The   Markov  matrix  $M$  encodes  the dynamics of the exclusion process:  
 the  non-diagonal  element  $ M_1(\mathcal{C},\mathcal{C}')$
 represents
 the  transition rate  from configuration $\mathcal{C}'$ to $\mathcal{C}$
 where  a particle hops  in the  forward ({\it i.e.}, anti-clockwise)
   direction,
  the  non-diagonal  element  $ M_{-1}(\mathcal{C},\mathcal{C}')$
 represents
 the  transition rate  from configuration $\mathcal{C}'$ to $\mathcal{C}$
where  a particle hops  in the  backward  ({\it i.e.}, clockwise)  direction.
 The diagonal term $M_0(\mathcal{C},\mathcal{C}) =
  -  \sum_{\mathcal{C}'\neq \mathcal{C}  }\left(  M_1(\mathcal{C}',\mathcal{C})
  +   M_{-1} (\mathcal{C}',\mathcal{C}) \right)   $ 
represents  the exit rate
 from  the configuration $\mathcal{C}$.

 \subsection{Generalized master equation for current statistics}

    We  call $Y_t$ the total distance
 covered by all the particles between time 0 and time $t$
 and   $P_t(\mathcal{C}, Y)$ the joint probability
 of being  in the configuration $\mathcal{C}$ at time $t$   
  with  $Y_t = Y$. An evolution equation,
 analogous to equation~(\ref{eq:Markov}), 
 can  be written for  $P_t(\mathcal{C}, Y)$ as follows~:
\begin{equation}
    \frac{d}{dt} P_t(\mathcal{C}, Y)  = \sum_{\mathcal{C}'} \Big(
   M_0(\mathcal{C},\mathcal{C}') P_t(\mathcal{C}', Y)  
+   M_1(\mathcal{C},\mathcal{C}')  P_t(\mathcal{C}', Y -1)  
+   M_{-1} (\mathcal{C},\mathcal{C}')  P_t(\mathcal{C}', Y +1)  
        \Big)   \, . 
 \label{eq:Markov2}
\end{equation}
  We now recall how the full statistics of $Y_t$ can be determined
  \cite{DLebowitz,lebowspohn}.
 In terms of   the generating function  $F_t(\mathcal{C})$ defined as
 \begin{equation}
  F_t(\mathcal{C}) =  \sum_{ Y =-\infty}^{+\infty}
  {\rm e}^{\gamma Y} P_t(\mathcal{C}, Y) \, ,
 \label{eq:defF}
\end{equation}
  equation~(\ref{eq:Markov2}) takes the simpler
 form~:
\begin{equation}
    \frac{d}{dt} F_t(\mathcal{C})  = \sum_{\mathcal{C}'} \Big(
   M_0(\mathcal{C},\mathcal{C}') 
+ {\rm e}^\gamma  M_1(\mathcal{C},\mathcal{C}')  
+   {\rm e}^{-\gamma}  M_{-1} (\mathcal{C},\mathcal{C}')   
        \Big)  F_t(\mathcal{C}') =  \sum_{\mathcal{C}'} 
  M(\gamma)(\mathcal{C},\mathcal{C}')   F_t(\mathcal{C}')
   \, . 
 \label{eq:Markov3}
\end{equation}
This equation is   similar to the original 
 Markov  equation~(\ref{eq:Markov}) for the probability
 distribution  $P_t(\mathcal{C})$ but where the original
 Markov matrix $M$ is deformed into $ M(\gamma)$  which is given by
 \begin{equation}
  M(\gamma) =   M_0 +  {\rm e}^\gamma  M_1 +  {\rm e}^{-\gamma}  M_{-1} \, . 
 \label{eq:defMgamma}
\end{equation}
 We emphasize  that  $M(\gamma)$,
  that   governs the evolution of  $F_t(\mathcal{C})$, 
  is not a Markov matrix for $\gamma \neq 0$
 (the sum of the elements in  a  given column does not vanish).

 \subsection{Long time limit and maximal eigenvalue}

 In the long time limit, $ t \to \infty$, 
  the behaviour of   $F_t(\mathcal{C})$
 is dominated by the largest eigenvalue $\lambda(\gamma)$ 
 of the matrix  $M(\gamma)$~:
  \begin{equation}
   F_t(\mathcal{C}) \to  {\rm e}^{E_{{\rm max}}(\gamma) t }
  \langle \, \mathcal{C} \,  | \,  E_{{\rm max}}(\gamma) \,  \rangle \,  , 
 \label{eq:limF}
\end{equation}
 where the ket $ |   E_{{\rm max}}(\gamma)   \rangle $ is the eigenvector
 corresponding to the  largest eigenvalue. 
 Therefore, when  $ t \to \infty$, we obtain  
 \begin{equation}
    \langle  \,  {\rm e}^{\gamma Y_t}\,  \rangle = 
  \sum_{\mathcal{C}}   F_t(\mathcal{C}) 
  \sim  {\rm e}^{E_{{\rm max}}(\gamma)t} 
   \, .
 \label{eq:limF2}
\end{equation}
 More precisely, we have
  \begin{equation}
   \lim_{t \to \infty} \frac{1}{t}
 \log   \langle  \,  {\rm e}^{\gamma Y_t}\,  \rangle = 
  E_{{\rm max}}(\gamma)   \, .
 \label{eq:limF3}
\end{equation}
The function $E_{{\rm max}}(\gamma)$  contains the complete
 information about the  cumulants  of the total current $Y_t$
 in the long time limit. For example, the total current $J$
 and the diffusion constant $\Delta$ are given by  
\begin{eqnarray}
 J =  \lim_{t \to \infty} \frac{ \langle Y_t \rangle}{t} &=&
 \frac{ {\rm d}  E_{{\rm max}}(\gamma)}{ {\rm d} \gamma} 
 \Big|_{\gamma =0} \, ,  \label{eq:defJ}  \\ 
 \Delta =  \lim_{t \to \infty} 
\frac{ \langle Y_t^2 \rangle - \langle Y_t \rangle^2 }{t}  &=&
  \frac{ {\rm d}^2  E_{{\rm max}}(\gamma)}{ {\rm d} \gamma^2} 
 \Big|_{\gamma =0}  \label{eq:defDelta}  \, .
 \end{eqnarray}
 Thus, the cumulants of $Y_t$ can be determined by carrying out 
 a perturbative expansion of  $E_{{\rm max}}(\gamma)$ with respect
 to $\gamma$ (a similar method has been used, in a different context,
  in  \cite{novotny}).
 The importance of the maximal eigenvalue $E_{{\rm max}}(\gamma)$ of 
 the matrix  $M(\gamma)$  also stems from  the fact that 
  it is closely related to the 
 large deviation function   $G$ for the total current. 
  We recall that the large deviation function $G$  is 
  defined as
\begin{equation}
 G(j) =  \lim_{t \to \infty} \frac{1}{t}
 \log \Big[ {\rm Prob}
  \left(      \frac{Y_t}{t} = j \right)   \Big] \, .
 \label{eq:defLDG}
  \end{equation}
   From equations~(\ref{eq:limF2}) and~(\ref{eq:defLDG}), we find 
  \begin{equation} 
 \langle  \,  {\rm e}^{\gamma Y_t}\, 
  \rangle \sim  {\rm e}^{E_{{\rm max}}(\gamma)t}
 \sim \int  {\rm e}^{ t(G(j) + \gamma j )}  {\rm d}j \, , 
\label{eq:LegT}
  \end{equation}
  and deduce by the saddle-point approximation   that the  maximal eigenvalue
 $E_{{\rm max}}(\gamma)$ is the Legendre transform of 
 the large deviation function  $G(j)$
  \begin{equation} 
    E_{{\rm max}}(\gamma) = \max_{j} \Big(  G(j) + \gamma j    \Big) \, .
\label{eq:LegT2}
  \end{equation}

\subsection{Restatement of the  problem}

 We want to study the statistical properties of the total
 current in the partially asymmetric exclusion process. We have seen
 that in the long time limit, the maximal eigenvalue 
  $E_{{\rm max}}(\gamma)$ of 
 the deformed  matrix  $M(\gamma)$ is 
 the generating function of the cumulants of the current, {\it i.e.},
 the power-series expansion of  $E_{{\rm max}}(\gamma)$ in the vicinity
 of $\gamma = 0$ allows us to determine  the  
 statistical properties  of the  current. 
 In the following sections, we shall first explain how 
 to  diagonalize   the Matrix   $M(\gamma)$  using the  Bethe Ansatz; 
 this method will allow us to write any 
  eigenvalue  of  $M(\gamma)$ as a  symmetric function
 of the roots of a  system of coupled polynomial equations
 (the Bethe  equations). Then,  we shall develop
 a perturbative scheme to  expand the   maximal eigenvalue 
  $E_{{\rm max}}(\gamma)$  in powers of 
 $\gamma$, when $\gamma \to 0$.  The first order expansion will
  give us the  the current $J$ and the second order
 term will lead to  the diffusion constant $\Delta$.

 \section{The Bethe equations}
\label{sec:Bethe}
 
The  deformed matrix   $M(\gamma)$ can be diagonalized by  Bethe
 Ansatz. A vector $P$ over the configuration space is an
 eigenvector of   $M(\gamma)$ if it satisfies 
 \begin{equation} 
  M(\gamma)  P = E(\gamma) P      \, . 
  \label{eq:eigenval}
 \end{equation}
 By representing a  configuration 
  by the positions of the $N$ particles
on the ring, $(r_1, r_2, \dots, r_N)$ with $1 \le r_1 < r_2 < \dots < r_N
\le L$,  the eigenvalue equation  (\ref{eq:eigenval})   becomes
\begin{eqnarray}
 &&  E P(r_1,\dots, r_N) =   \nonumber \\   &&
  \sum_i  p \left[  {\rm e}^{\gamma} 
             P(r_1, \dots, r_{i-1},\ r_i-1,\ r_{i+1}, \dots, r_N) 
 - P(r_1,\dots, r_n) \right] + 
    \nonumber \\      
  &&  \sum_j  q
 \left[ {\rm e}^{-\gamma}  P(r_1, \dots, r_{j-1},\ r_j+1,\ r_{j+1}, \dots, r_N)
              - P(r_1,\dots, r_N) \right] \, , 
\end{eqnarray}
where the sum runs over the indices  $i$ such that $ r_{i-1} < r_i-1$
 and   over the indices  $j$ such that $  r_j + 1 < r_{j+1} \,;$   these
 conditions  ensure  that the corresponding jumps  are  allowed.
  Following the coordinate  {\em Bethe Ansatz}, we assume that the 
  eigenvector  $P$ can be written in the form
\begin{equation}
  P(r_1,\dots,r_n) = \sum_{\sigma \in \Sigma_n} A_{\sigma}  \,  
         z_{\sigma(1)}^{r_1} \,  z_{\sigma(2)}^{r_2} \dots z_{\sigma(n)}^{r_n}
  \label{eq:ba} \, , 
\end{equation}
where $\Sigma_n$ is the group of the $n!$ permutations of $n$ indices. The
coefficients $\{A_{\sigma}\}$ are rational
 functions of the  fugacities $\{z_1, \dots, z_n\}$. 
 The  expression~(\ref{eq:ba}) represents an eigenvector of 
  $M(\gamma)$  if    $\{z_1, \dots, z_n\}$ 
    satisfy  the   {\it  Bethe equations} \cite{gwaspohn,ogkmrev}:
\begin{equation}
   z_i^L  = (-1)^{N-1} \prod_{j =1}^N
   \frac{q {\rm e}^{-\gamma}  z_i z_j - (p+q) z_i + p{\rm e}^{\gamma}  }
  {q{\rm e}^{-\gamma}  z_i z_j -(p+q) z_j +p{\rm e}^{\gamma} } \,\,\,
  {\rm  for }\,\,\,\,  i =1 \ldots N\, , 
\label{eq:Bethelambda}
\end{equation}
and the corresponding eigenvalue of $M(\gamma)$ is given by
\begin{equation}
 E(\gamma; z_1, z_2 \ldots  z_N ) = p {\rm e}^{\gamma}  \sum_{i =1}^N 
   \frac{1}{ z_i}  + 
        q {\rm e}^{-\gamma}  \sum_{i =1}^N   z_i \, - N (p+q)  \, .
\label{eq:vplambda}
\end{equation}
  For   $\gamma = 0$, we know that
 the maximal eigenvalue of the Markov matrix $M$ is equal to 0
 and corresponds to the degenerate solution $z_i =1$ for all $i$. 

 \hfill\break
{\it Remark: The  Gallavotti-Cohen Invariance.} 
  The Bethe  equations~(\ref{eq:Bethelambda}) and
 equation~(\ref{eq:vplambda}) are invariant under the transformation
  $               z  \rightarrow \frac{1}{z}  \, ,$ 
               $  \gamma  \rightarrow  \log\frac{q}{p}- \gamma   .$
 This symmetry implies that the spectrum of  $M(\gamma)$
 and that of  $M( \log\frac{q}{p} -\gamma)$  are identical.
 This functional identity  is  satisfied 
 in particular  by the largest eigenvalue
 of $M$ and we have
$ E_{{\rm max}}(\gamma) =  E_{{\rm max}}( \log\frac{q}{p}- \gamma) \,.$
 This  identity  implies, using 
 equation~(\ref{eq:LegT2}), that  the large deviation function 
  satifies the symmetry 
\begin{equation}
 G(j) =   G(-j) -  \left(\log\frac{q}{p}\right) j  \, .
 \label{eq:GC2}
  \end{equation}
This relation is a special case of the general {\it Fluctuation
 Theorem} valid for  a large class of
 systems far from equilibrium. It was derived
 for more  general Markovian systems in \cite{lebowspohn}.

\subsection{A useful change of variables}
 We introduce  $N$  auxiliary variables $(y_1,\ldots ,y_N)$ defined as 
\begin{equation} 
   y_i =  \frac{ 1 - {\rm e}^{-\gamma} z_i} { 1 - x {\rm e}^{-\gamma} z_i}
 {\hskip 1 cm}   {\rm  for }\,\,\,\,  i =1 \ldots N\, , 
  \label{eq:defyi}
  \end{equation}
where we have introduced the {asymmetry parameter} $x$:
 \begin{equation} 
 x = \frac{q}{p} \, .
 \label{eq:defx}
 \end{equation}
 We remark that the change of variables~(\ref{eq:defyi}) is ill-defined
 for $x =1$ which corresponds to the symmetric exclusion process.
 In the following, our calculations will always be performed
 for $x < 1$. Our results will extend to the  symmetric case   by taking 
  the limit $ x \to 1$ in the final expressions. 
     The Bethe equations~(\ref{eq:Bethelambda}) now  become
\begin{equation}
     {\rm e}^{L\gamma} \left( \frac{ 1-y_i}{1- x y_i} \right)^L =
      - \prod_{j =1}^N \frac{ y_i -  x y_j }{x y_i - y_j }
 \,\,\,\,   {\rm  for }\,\,\,\,  i =1 \ldots N\, . 
\label{eq:BetheEq}
\end{equation}
These equations are simpler than the original ones because they involve
  only  linear polynomials in   the $y_i$'s.
 By taking the  product of the Bethe equations~(\ref{eq:BetheEq})
 over all the values of $i$, we obtain 
 \begin{equation}
   \left( {\rm e}^{N\gamma}\prod_{i =1}^N \frac{ 1-y_i}{1- x y_i} \right)^L =
  (-1)^N  \prod_{i,j =1}^N \frac{ y_i -  x y_j }{x y_i - y_j }
 = (-1)^{N+N^2} = 1 \, .
\label{momentum}
\end{equation}
 This relation stems from the translation invariance of the model
 (momentum conservation).

 In terms of the $y_i$'s,  the eigenvalue~(\ref{eq:vplambda})  reads 
\begin{equation} 
 E(\gamma) = p(1-x)   \sum_{i =1}^N  \left(   \frac{1}{ 1-y_i} - 
  \frac{1} {1- x y_i}  \right)  \, .
\label{eq:valeurpropre}
\end{equation}
 When  $\gamma  \to  0$,   all the  roots 
 $y_i(\gamma)$ that correspond to the maximal eigenvalue 
 $E_{{\rm max}}(\gamma)$     of  $M(\gamma)$
  converge to the degenerate solution 
 $ \lim_{\gamma \to 0} y_i = 0 $ 
 and the maximal eigenvalue of  $M(\gamma)$
 also  converges to 0. Using equation~(\ref{momentum}), 
  we  therefore  find  that,  for small enough values of $\gamma$,  
   the   roots  $y_i(\gamma)$  satisfy the relation 
 \begin{equation}
    {\rm e}^{N\gamma} \prod_{i =1}^N \frac{ 1-y_i}{1- x y_i}   = 1 \, . 
 \label{eq:Bethesupplem}
 \end{equation}
  This relation, which is a simple consequence of the Bethe equations,
 will be useful in the following  to select the Bethe roots that correspond
 to  $E_{{\rm max}}(\gamma)$. 

\subsection{The TASEP case}

 The Bethe equations~(\ref{eq:BetheEq}) are a  coupled non-linear
  system of polynomial equations in the variables $y_1,\ldots,y_N$. Deriving 
 exact results from  these equations is a daunting task. However,  
 for the special case of  the
 totally asymmetric exclusion process (TASEP), which corresponds to $p =1$ and
  $q =x =0$, the  Bethe equations  can be  reduced  to 
  an effective one variable  problem.  Indeed, for $x=0$, the equations
 (\ref{eq:BetheEq}) read 
 \begin{equation} 
  {\rm e}^{L\gamma} (1-y_i)^L =
  (-1)^{N-1}\frac{ y_i^N }{\prod_{j =1}^N y_j } \, . 
\end{equation}
  Thus, all the Bethe roots $y_i$ are solutions of the 
 one-variable polynomial equation 
\begin{equation} 
    {\rm e}^{L\gamma} (1-T)^L +  C T^N = 0 \, ,
 \end{equation}
 where the constant $C$ must be determined self-consistently
  by  the relation
\begin{equation} 
  C = (-1)^{N}\prod_{j =1}^N\frac{1}{ y_j } \, . 
 \end{equation}
 This crucial  `decoupling'
  property of the the Bethe equations for $x =0$, has lead 
 to  an exact calculation of  the TASEP spectral gap
 \cite{dhar,gwaspohn,ogkm1} and has allowed 
   Derrida and Lebowitz  to 
  calculate   the complete  large deviation function
 of the  current  for any  finite values of $L$
 and $N$ \cite{DLebowitz}.  This effective decoupling
  also explains 
 the spectral degeneracies of the  TASEP Markov matrix \cite{ogkm2}.
   Hence, the use of the  Bethe Ansatz has been restricted mostly to
    TASEP  (for a review see e.g. \cite{ogkmrev}).

  For the partially asymmetric exclusion
 process, the   Bethe equations are highly coupled to one another 
 and  can not be simply  reduced  
 to an effective  one variable equation.  Because of
 this technical  difficulty  for  $0 < x < 1$, 
 it has not been possible to extract from the Bethe Ansatz  any 
  exact solution  for finite systems.  
  However, when    $L \to \infty$,  the Bethe equations reduce
 to an integro-differential equation for the density of roots, which
 was  analyzed by Kim et al.~\cite{kim1,kim2}
 to derive the spectral gap and the current  large deviation function.

\section{Reformulation of the   Bethe Equations}
 \label{sec:FBA}
 
 We note  that in the $N$  Bethe equations~(\ref{eq:BetheEq})
 all the variables $y_i$ play a similar role. This remark  suggests
 that we should  introduce an auxiliary variable $T$  that plays a symmetric
 role with respect to all the  $y_i$'s. We suppose that 
$T$ satisfies the following equation
\begin{equation}
     {\rm e}^{L\gamma} \left( \frac{ 1 -  \, \, T}{1- x T} \right)^L =
      - \prod_{j =1}^N \frac{  \, \, T   \,  -  x y_j }{x T - y_j }
 \,\,\,\,   {\rm  for }\,\,\,\,  i =1 \ldots N\, ,  
\label{variablesuppl}
\end{equation}
 where the $y_i$'s are now interpreted as {\it parameters} of the 
 problem. This expression  can be rewritten as a  one variable 
 polynomial equation for the unknown  $T$:
\begin{equation}
   P(T) = 0 \,\,\,\,
 \hbox{with } \,\,\, 
  P(T) = {\rm e}^{L\gamma} (1 -   T)^L \prod_{j =1}^N (x T - y_j)
 + (1 - x T)^L \prod_{j =1}^N  (T -  x y_j)  \, . 
\label{eq:defP}
\end{equation}
 The $N$  Bethe equations~(\ref{eq:BetheEq}) imply  that 
    $y_i$ is a root of  $P(T)$ for  $i=1,\ldots,N.$ 
 Thanks to the  auxiliary variable $T$, the 
  Bethe equations have    been reduced to an effective
 one variable problem with $N$ parameters. We can now proceed as follows:
 (i) Find the roots of the polynomial  $P(T)$ 
 with  the  unknown $T$  and with  $N$ parameters  $y_1,\ldots,y_N$.
 (ii) Select $N$ roots, amongst the $L+N$  solutions of  $P(T)=0$, 
 and identify these selected roots to the   $y_i$'s. This identification  
  leads to $N$ self-consistent  equations (recall that  for TASEP
 we had only one  self-consistency condition).

 It is possible to perform these steps
  using contour integration
 in the complex plane as in the TASEP 
 case \cite{appert,DLebowitz,evans,ogkm1}. However, the  calculations
 will be  greatly simplified if the 
  problem  is  formulated in  a purely algebraic manner, as follows. 
   Let us  define the polynomial
 $Q(T)$ as
  \begin{equation}
    Q(T) =  \prod_{j =1}^N  (T -   y_j)  \, . 
\label{eq:defQ}
\end{equation}
 The roots of $Q$ are exactly the Bethe roots $y_1,\ldots,y_N$ 
 (equivalently,  $Q$  is the generating function of the 
 symmetric polynomials in $y_1,\ldots,y_N$). 
 The polynomial   $P(T)$, defined in equation~(\ref{eq:defP}),
  can then be written  as follows
\begin{equation}
     P(T) = {\rm e}^{L\gamma} (1 -   T)^L  Q(xT) +
    (1 -   xT)^L x^N  Q\left(\frac{T}{x}\right)  \, . 
\label{eq:exprP}
\end{equation}
 The fact that the Bethe roots $y_1,\ldots,y_N$  are roots
 of the  polynomial   $P(T)$ implies that  $Q(T)$  {\it divides} $P(T).$
 Therefore there exists a  polynomial   $R(T)$ of degree $L$
 such that   $P(T) = Q(T) R(T)$, {\it i.e.}, such that
 \begin{equation}
   Q(T) R(T) =  {\rm e}^{L\gamma} (1 -   T)^L  Q(xT) +
    (1 -   xT)^L x^N  Q\left(\frac{T}{x}\right)  \, . 
\label{FBA}
\end{equation}

  Substituting  $T = y_i$ in this equation and  taking into account that $y_i$
 is a root of $Q(T)$ we obtain
\begin{equation}
 {\rm e}^{L\gamma} \left( \frac{1 -   y_i}{1 - xy_i}\right) ^L
 = -  x^N \frac{  Q\left(\frac{y_i}{x}\right) }{ Q(xy_i)}   \, .
 \end{equation}
 Using the expression~(\ref{eq:defQ}) for $Q(T)$, 
 we find  that  this relation is identical to
 the  Bethe  equation~(\ref{eq:BetheEq}). 
 We remark that this  reformulation of the Bethe equations as a  problem 
  of polynomial divisibility  has been used in various
  contexts  \cite{baxter,brunet,razumov} and is closely related
 to  the functional Bethe  Ansatz \cite{baxter,razumov,babelon}. 

\subsection{Expression of the eigenvalue}

 The eigenvalue  $E(\gamma)$, defined in
  equation~(\ref{eq:valeurpropre}),   can  be expressed 
 in terms of  the polynomial  $Q(T)$:
\begin{equation} 
 E(\gamma) = p(1-x)     \left(   \frac{Q'(1)}{Q(1)} - 
  \frac{1}{x} \frac{Q'(1/x)}{Q(1/x)}  \right)  \, .
\label{eq:vp2}
\end{equation}
  This formula can  be simplified  with the help of the 
 `Q-R equation'~(\ref{FBA})  as follows.
   Substituting  $T=1$ in the  equation~(\ref{FBA}) we find
 \begin{equation} 
   Q(1) R(1) = (1 -   x)^L x^N  Q\left(\frac{1}{x}\right) \, .
\label{QR1}
\end{equation}
   If we differentiate equation~(\ref{FBA}) with respect to $T$ and then
 substitute $T=1$ we obtain
  \begin{equation}
  Q'(1) R(1) + Q(1) R'(1) = -Lx^{N+1} (1 -   x)^{L-1}Q\left(\frac{1}{x}\right)
   + x^{N-1}(1 -   x)^L  Q'\left(\frac{1}{x}\right) \, .
\label{primeQR1}
\end{equation}
  Taking the ratio  of the last two equations, we find that 
$E(\gamma)$ can be rewritten as 
\begin{equation} 
  \frac{ E(\gamma)}{ p(1-x)} = - \frac{Lx}{1-x} - \frac{ R'(1)}{ R(1)} \, . 
\label{eq:vp3}
\end{equation}
This is the  expression of  $E(\gamma)$  that  will be used  in the sequel.

   Equation~(\ref{eq:Bethesupplem}), that allows
 to select the  roots $y_i$  corresponding to the maximal eigenvalue,
  is similarly  rewritten in terms of   $Q$ and $R$  as  follows
\begin{eqnarray}
   {\rm e}^{N\gamma} \frac{Q(1)}{ x^N  Q\left(\frac{1}{x}\right) } = 1  \, .
 \label{eq:auxil1}
\end{eqnarray}
 Using equation~(\ref{QR1}), an  alternative form is obtained  
  \begin{eqnarray}
    R(1) =  {\rm e}^{N\gamma} (1 -   x)^L  \, .
 \label{eq:auxil2}
\end{eqnarray}
 This   relation will be very useful in the sequel    to simplify some
 calculations.

\section{Perturbative solution of the Functional Bethe Ansatz equations}
\label{sec:pert}

 In  this  section,  we explain
 how to solve equation~(\ref{FBA}) order by order  in $\gamma$
  for the roots $y_i$ that correspond  to the maximal eigenvalue
 of the matrix $M(\gamma)$.

  We first  develop  the polynomials $Q$ and $R$
in powers of  $\gamma$
\begin{eqnarray}
      Q(T) =  \prod_{j =1}^N  (T -   y_j) &=& \sum_{n=0}^\infty \gamma^n Q_n(T)
 =  Q_0(T) + \gamma  Q_1(T) + \gamma^2  Q_2(T) + \ldots 
  \label{eq:devQ}  \\
         R(T)  &=& \sum_{n=0}^\infty \gamma^n R_n(T)
 =  R_0(T) + \gamma  R_1(T) + \gamma^2  R_2(T) + \ldots 
   \, .
   \label{eq:devR}
\end{eqnarray}
 We note that the degree of the polynomials  $Q_n(T)$ for $n \ge 1$ is
 at most $N-1$. 
 For $\gamma = 0$, we know that  $E_0 = 0$ 
 and that this  maximal eigenvalue is obtained
 for $y_i =0$. Therefore, we have
\begin{eqnarray}
 Q_0(T) =  T^N \, \,\,\,\,\,\,  \hbox{ and }
 {\hskip 1cm}   R_0(T) =  (1 -   xT)^L +  x^N (1 -   T)^L   \, . 
  \label{eq:ordre0}
\end{eqnarray}

 By substituting the power series~(\ref{eq:devQ}) and~(\ref{eq:devR})
 in the QR-equation~(\ref{FBA}), we obtain a  hierarchical system of linear
 equations  for the polynomials  $R_n(T)$ and   $Q_n(T)$. This system
  can be solved order by order  by using the known `initial conditions' 
 $Q_0(T)$  and  $R_0(T)$.

 We now solve the QR-equation to the first and second orders.

 \subsection{First order calculation}

  At first order, the QR-equation~(\ref{FBA}) becomes 
\begin{eqnarray}
 Q_1(T)  \left[ (1 -   xT)^L  +  x^N (1 -   T)^L   \right]
      + T^N R_1(T) =  (1 -   T)^L  Q_1(xT) +
    (1 -   xT)^L x^N  Q_1\left(\frac{T}{x}\right) 
   + L x^N  (1 -   T)^L  T^N \, , 
  \label{eq:ordre1}
\end{eqnarray}
  and the  auxiliary equation~(\ref{eq:auxil1})  becomes 
\begin{equation}
     Q_1(1) - x^N Q_1\left(\frac{1}{x}\right) = -N  \, .
\label{eq:auxilordre1}
\end{equation} 
   It is simpler to define the  polynomial
 \begin{equation}
 B_1(T) =  Q_1(T) -  x^N  Q_1\left(\frac{T}{x}\right) \, ,
 \label{eq:defB1}
\end{equation}
  and to  rewrite   equations~(\ref{eq:ordre1}) and~(\ref{eq:auxilordre1})
     as follows
\begin{eqnarray}
  (1 -   xT)^L B_1(T) - (1 -   T)^L B_1(xT) &=& 
  T^N \left(    L x^N  (1 -   T)^L -     R_1(T)     \right) \, .
    \label{eq:B1}    \\
    B_1(1) &=& - N   \, .  \label{eq:auxB1} 
\end{eqnarray}
  Because    $B_1(T)$ and   $Q_1(T)$ are 
  of degree $\le N-1$ and noting that  the term  on 
  the  r.h.s. of equation~(\ref{eq:B1}) is divisible by $T^N$,
  we can reduce  this equation modulo $T^N$ and write
 \begin{equation}
  (1 -   xT)^L B_1(T) - (1 -   T)^L B_1(xT) \equiv 0  \,\,\, [T^N] \, .
  \label{eq:B1modTN}
\end{equation}
This equation allows to determine the polynomial $B_1(T)$ up to
 a multiplicative constant $\beta_0$
\begin{equation}
  B_1(T) \equiv  \beta_0 (1 -   T)^L  \,\,\,\,  [T^N] ,  \,\,\, {i.e.}, \,\,\,
  B_1(T) = \beta_0 \sum_{k=0}^{N-1} (-1)^k
  \left(\begin{array}{c} L \\ k \end{array} \right) T^k \,.
\label{eq:solB1}
\end{equation}
 The constant $\beta_0$ is fixed using equation~(\ref{eq:auxB1}).
 Using the binomial identity~(\ref{Bin3}), we find 
 \begin{equation}
  -N  =  \beta_0 \sum_{k=0}^{N-1} (-1)^k
  \left(\begin{array}{c} L \\ k \end{array} \right) = 
  \beta_0 (-1)^{N-1}\left(\begin{array}{c} L-1\\ N-1 \end{array} \right)\, ,
  {\hskip 0.7 cm }  i.e.,  {\hskip 0.4 cm }
\beta_0 =  \frac{ (-1)^{N} L}
 {\left(\begin{array}{c} L\\ N \end{array} \right)}\, . \label{eq:beta0}
\end{equation}
   From this relation it follows  that
 \begin{equation}
 Q_1(T) = \sum_{k=0}^{N-1} q_k^{(1)}T^k  \,\,\,\,
 \hbox{ with } 
   q_k^{(1)}  = 
  \frac{ (-1)^{N+k} L}{\left(\begin{array}{c} L\\ N \end{array} \right)}
  \frac{\left(\begin{array}{c} L\\ k \end{array} \right)}
     { 1 - x^{N-k}}  \,. 
\label{solQ1}
\end{equation}
 
  Using  this formula and equation~(\ref{eq:B1}) the following exact 
 expression  for $R_1(T)$ is  obtained:
 \begin{equation}
    R_1(T) =  L x^N  (1 -   T)^L +  (-1)^{N}
  \frac{L}{\left(\begin{array}{c} L\\ N \end{array} \right)}
 \sum_{p=0}^{N-1} \sum_{r=0}^{L}(-1)^{p+r} 
 \left(\begin{array}{c} L\\ p \end{array} \right)
  \left(\begin{array}{c} L\\ r \end{array} \right) (x^p - x^r) T^{p+r -N} \, .
 \label{solR1}
\end{equation}
 All negative powers of T   in  the  above  expression cancel out
 for the following reason:  the coefficient of
  a term  of the type $T^{-d}$ with $d >0$ is obtained by imposing
 the condition $p+r = N-d$ to 
  the double sum  in equation~(\ref{solR1}).
  Because of this condition,  the indices   $p$ and $r$
 can  vary  only from  $0$ to $N-d$  and they 
 both  have  the same effective range. 
 The sum in  equation~(\ref{solR1}) is 
  antisymmetric  with respect to  $p$ and $r$ and therefore it vanishes.
   This proves that   $R_1(T)$ is indeed a polynomial.

\subsection{Second  order calculation}

   At second order, the polynomial $B_2(T)$   defined as 
\begin{equation}
 B_2(T) =  Q_2(T) -  x^N  Q_2\left(\frac{T}{x}\right) \, ,
 \label{eq:defB2}
  \end{equation}
satisfies the following equation 
\begin{equation}
  (1 -   xT)^L B_2(T) - (1 -   T)^L B_2(xT)  = 
    L (1 -   T)^L  Q_1(xT) - R_1(T) Q_1(T)  +
 T^N\left( x^N \frac{L^2}{2} (1 -   T)^L -  R_2(T) \right) \, . 
 \label{eqB2}
\end{equation}
 If we write this relation  modulo $ T^N $ we obtain the simpler equation 
\begin{equation}
  (1 -   xT)^L B_2(T) - (1 -   T)^L B_2(xT)  \equiv
      L (1 -   T)^L  Q_1(xT) - R_1(T) Q_1(T)   \,\,\, [T^N] \, ,
 \label{eq:defB2modTN}
\end{equation}
  where  the expressions for  $Q_1(T)$  and  $R_1(T)$ are given in    
 equations~(\ref{solQ1}) and ~(\ref{solR1}) respectively.
  At order 2,  the  auxiliary equation~(\ref{eq:auxil1})  becomes 
\begin{equation}
   B_2(1) =  Q_2(1) - x^N Q_2\left(\frac{1}{x}\right) = 
 -NQ_1(1) -\frac{N^2}{2}  \, .
\label{eq:auxilordre2}
\end{equation} 
 The polynomial $B_2(T)$ is    the sum of a 
 special solution  $\tilde{B_2}(T)$  of  equation~(\ref{eq:defB2modTN}) and
 of a term that is  proportional to  $B_1(T)$,  the solution of the
 homogeneous equation~(\ref{eq:B1modTN}), {\it i.e.,} 
\begin{equation}
  B_2(T) = \tilde{B_2}(T) + C B_1(T) \, . 
 \label{eq:exprB2}
 \end{equation}
  The  proportionality
 constant  $C$ is fixed by using  the auxiliary
  equation~(\ref{eq:auxilordre2}), which   leads to 
\begin{equation}
   C = \frac{\tilde{B_2}(1)}{N} + Q_1(1) + \frac{N}{2}  \, , 
 \label{eq:solC}
 \end{equation}
 where  we have used    $B_1(1) = -N$  from equation~(\ref{eq:auxB1}). 

 A special  solution to the polynomial equation~(\ref{eq:defB2modTN}) 
  is given by  
\begin{eqnarray}
    \tilde{B_2}(T) &=&  \sum_{k=0}^{N-1}(1 - x^{N-k}) q_k^{(2)}T^k \, ,
 \label{def:tildeB2}  \\ 
 \hbox{ with }   q_k^{(2)}  &=& 
  \frac{ (-1)^{N+k+1} L^2}{\left(\begin{array}{c} L\\ N \end{array} \right)^2}
\frac{1} { 1 - x^{N-k}} \left\{
 \sum_{r=1}^{N-1} 
 \frac{     \left(\begin{array}{c} L\\ N + r \end{array} \right)
   \left(\begin{array}{c} L\\ k- r \end{array} \right)  x^{r} }
     { 1 - x^{r} } + 
    \sum_{r=0}^{N-1} 
  \frac{\left(\begin{array}{c} L\\ N + r \end{array} \right)
   \left(\begin{array}{c} L\\ k- r \end{array} \right) }
     { 1 - x^{N+r -k}}   
    \right\}   \, .
\label{eq:solB2}
\end{eqnarray}
 The main  steps to derive   equation~(\ref{eq:solB2})
   are given in Appendix~\ref{AppendixSOLQ2}.
 Finally, 
 the polynomial $Q_2(T)$ is given by the linear combination
 \begin{equation}
 Q_2(T) = \sum_{k=0}^{N-1} q_k^{(2)} T^k   + 
  C  \sum_{k=0}^{N-1} q_k^{(1)} T^k \, , 
\label{solQ2}
\end{equation}
  where  the constant $C$ is  given in equation~(\ref{eq:solC}).

\section{Exact Formulae for the mean current and its fluctuations}
\label{sec:formules}

Solving the Q-R  equation allows us to calculate the expansion of
 the largest eigenvalue  $E_{{\rm max}}(\gamma)$,  order by order, 
 and to calculate the cumulants  of the total current.
  The largest eigenvalue  $E_{{\rm max}}(\gamma)$,
 can  be expanded with respect to the parameter $\gamma$ as follows 
 \begin{equation} 
 E_{{\rm max}}(\gamma) = p(1-x)   \sum_{n=0}^\infty \gamma^n E_n \, .
\label{eq:devvp}
\end{equation}
 Using  equations~(\ref{eq:vp3}),~(\ref{eq:auxil2}) and~(\ref{eq:devR}), 
  the expansion of  $E_{{\rm max}}(\gamma)$ is  given by
\begin{equation} 
 \frac{E_{{\rm max}}(\gamma)}{p(1-x)} = -\frac{Lx}{1-x}
  -\frac{R_0'(1)}{(1-x)^L} 
 + \gamma 
\left( \frac{N R_0'(1)}{(1-x)^L} -\frac{ R_1'(1)}{(1-x)^L} \right)
 + \gamma^2
 \left(-\frac{N^2 R_0'(1)}{2(1-x)^L} + \frac{N R_1'(1)}{(1-x)^L}
         - \frac{R_2'(1)}{(1-x)^L} \right) +\ldots
 \label{devpEmax}
\end{equation}
  From the expression~(\ref{eq:ordre0}) for  $R_0(T)$,
 we find  that
\begin{equation} 
  \frac{R_0'(1)}{(1-x)^L}  =  -\frac{Lx}{1-x} \, ,
\label{valR0p1}
\end{equation}
 and  we verify that the zeroth-order term  $E_0$ 
  in  $E_{{\rm max}}(\gamma)$ vanishes.

\subsection{Calculation of the Current}

 The  current $J$, defined in  equation~(\ref{eq:defJ}),
 corresponds to the coefficient of $\gamma$
  in  the expansion of  $E_{{\rm max}}(\gamma)$. 
  To  determine   $R_1'(1)$,   we
  start with equation~(\ref{primeQR1})   and  expand
 it to the first order in $\gamma$:
\begin{equation} 
 \frac{ R_1'(1)}{(1-x)^L} = -N^2 + 
 \frac{L  x}{1-x}\left( Q_1(1) - x^N  Q_1\left(\frac{1}{x}\right)  \right)
  - Q_1'(1) +  x^{N-1}  Q_1'\left(\frac{1}{x}\right)
 =  -N^2 +  \frac{L  x}{1-x} B_1(1) -  B_1'(1) \, ,
\end{equation}
 where in the last equality we have used the definition of  $B_1(T)$ as
 given by equation~(\ref{eq:defB1}). We know that $B_1(1) = -N$
 from equation~(\ref{eq:auxB1});    the
 value of $ B_1'(1)$ is readily obtained  from the  expression
 of $B_1(T)$ given  in equations~(\ref{eq:solB1}) and~(\ref{eq:beta0}): 
\begin{equation} 
 B_1'(1) = -LN \frac{N-1}{L-1}   \, .
 \label{Bp1of1}
\end{equation}
 Thus, we have
 \begin{equation} 
 \frac{ R_1'(1)}{(1-x)^L} = -N \left(  \frac{Lx}{1-x} +
  \frac{L-N}{L-1}  \right) \, .
 \label{valR1p1}
\end{equation}
 Substituting this  expression in the coefficient of $\gamma$
 in equation~(\ref{devpEmax}), we find  that  the total current
 is given by 
\begin{equation} 
 J = p(1-x) \frac{N(L-N)}{L-1} \, . 
\label{eq:valcourant}
\end{equation}
This value  agrees,  of course,  with the known formula, which  is
 obtained  very simply by using the fact that all the stationary
 configurations of ASEP on a  ring are equiprobable. 
 We recall that $J$ represents the total current in the system;
 the current through a bond is given by $J/L$.
 Using  the Bethe Ansatz  to find  $J$  is certainly a very 
  complicated and  distorted   way
 to retrive a back-of-an-envelope calculation. However, $J$ is   
 one of the  simplest quantity associated with ASEP and 
 the fact that nobody  could  extract such an elementary
  formula  from   the Bethe equations
 has been  a standing  puzzle for a long time.  

\subsection{Calculation of the Diffusion constant}

 The second order term in the perturbative expansion~(\ref{devpEmax}) 
 allows us to calculate the diffusion constant. Indeed, thanks
 to equation~(\ref{eq:defDelta}), we find that $\Delta =2p(1-x) E_2$.
 Therefore, we have
\begin{equation}
 \Delta = 2p(1-x) \left(-\frac{N^2 R_0'(1)}{2(1-x)^L} + \frac{N R_1'(1)}{(1-x)^L}
         - \frac{R_2'(1)}{(1-x)^L} \right) \, .
 \label{DeltaR2}
\end{equation}
 Hence, in  order to calculate $\Delta$, we  also need  
   $R_2'(1)$, which is determined  in  Appendix~\ref{AppendixUsteps}.
 After gathering all relevant terms, we 
 are  finally  lead  to the exact formula
 for the diffusion constant of the total current for the 
 partially asymmetric exclusion process on a ring:
\begin{equation} 
 \Delta = \frac{2p(1-x)L}{(L-1)\left(\begin{array}{c} L\\ N \end{array}\right)^2 }
   \sum_{r=1}^{N} r^2 \frac{ 1 + x^r}{1 - x^r} 
 \left(\begin{array}{c} L\\ N +r  \end{array}\right) 
  \left(\begin{array}{c} L\\ N -r \end{array}\right)  \, . 
 \label{formuleCstedeDiff}
\end{equation}
 This formula agrees, of course, with the one obtained using
 the Matrix Representation method \cite{DMal}
 (in that work,
 the  fluctuations  of the current through a bond
 were calculated exactly {\it i.e.},  $\Delta/L^2$).
  From this exact
 expression, it is possible to deduce by finite size scaling
 that  a tagged particle in an infinite system  exhibits an anomalous
 diffusive behaviour with exponent $1/3$ (instead of one 1/2).
 By taking the continuous limit  $ L \to \infty$
  of equation~(\ref{formuleCstedeDiff})
 in the weakly asymmetric regime $x \to 1$, with scaling variable
  $\phi = (1-x)\sqrt{L}$, it is possible to derive a scaling
 function for the KPZ equation that describes the cross-over
 from the linear Edwards-Wilkinson regime to the non-linear
 KPZ regime. We refer for more details to \cite{DMal}.

 We emphasize that 
 the calculation of  $\Delta$ with the  Bethe Ansatz is of the
 same order of complexity as with 
  the Matrix method \cite{DMal} but it is  much simpler mathematically.
    The  Bethe Ansatz  requires only elementary
 mathematical objects such as polynomials and 
 involves systematic calculations,
 whereas for  the Matrix Ansatz one has to find (guess) a suitable algebra,
 prove that this  algebra solves the problem and then evaluate
  traces of various operators 
 requiring  the use
 of remarkable identities on q-binomials \cite{DMal}.

   Furthermore, 
 to calculate the higher  cumulants of the current, one has to solve 
 the QR-equation~(\ref{FBA}) to the
 suitable  order  in $\gamma$. By contrast,  there is absolutely no clue on 
 how to extend the  Matrix method to calculate, for example, 
 the third  cumulant of the current: the  form of the algebra  involved
 (if such an algebra does exist) is totally unknown. 
 
\section{Conclusion}

  Most of the analytical studies  of  the  ASEP are based on 
   two different  techniques, 
   the Matrix Product method and the Bethe Ansatz.  
 The Matrix representation  is suitable  to calculate
 stationary state observables, such as correlations, phase diagrams
  etc...  A major drawback of this method is that there
 is no constructive method to generate matrices that are suitable
 for a given stochastic model: one has to  rely   on educated 
 guesses, after some  trials and errors. Nevertheless,
 the   Matrix method,  when  applicable,    is  efficient and allows to
 derive elegant combinatorial results for finite systems.
  On the contrary, the Bethe Ansatz is a systematic procedure
 with such  a wide  range of applicability  that it  has grown into
 a  subfield of theoretical physics: the theory
 of integrable systems. There exists a 
  priori conditions, such as the Yang-Baxter relation, 
 that insure that a system is integrable ({\it i.e.},
  it can be analyzed by Bethe Ansatz).  Many methods
 have been developed to cope with the Bethe equations \cite{baxter,babelon}.
 However,  it is  very difficult to extract information for finite
 systems from the  Bethe equations and usually one has to analyze
 these equations in the thermodynamic limit. 
 
   For the TASEP, the  Bethe equations  have a fundamental 
   decoupling property
 that has lead to many exact results \cite{gwaspohn, ogkm1,ogkm2}
 and in particular to the calculation of an exact formula
 for the large deviation function \cite{DLebowitz,appert}. 
 For the partially asymmetric case,  the Bethe equations are strongly
 coupled and therefore  they have been rarely  used. The only 
 exact results derived from them were obtained  by Kim et al.
 in the  limit of an infinite size \cite{kim1,kim2}. In this paper, 
  we have been able to overcome this  technical difficulty
 thanks to a  reformulation of the Bethe equations   as a mere 
  problem of   polynomial divisibility  that can be solved perturbatively
 in the fugacity parameter. 
   We have calculated the mean value $J$ 
  of the current and its fluctuations $\Delta$. Obviously,  the calculation
 of $J$ from Bethe Ansatz is much more difficult than the elementary
 derivation. However,  
 the calculation of  $\Delta$ with the  Bethe Ansatz is  less 
   difficult  than that with    the Matrix method \cite{DMal}. 
   Furthermore,  the  perturbative analysis  of the 
  Bethe Ansatz can be extended a priori to any order 
  to  derive   higher  cumulants of the current. It is not 
 known if the  Matrix method can be applied to such calculations.

   The reformulation of the  Bethe equations that we used  here, 
  is akin to the functional Bethe Ansatz \cite{baxter,babelon, razumov}.
  This method can be generalized to many other problems:
   higher moments of the ASEP current (S. Prolhac, in preparation),
  subleading  correction to the large deviation
    function of the symmetric exclusion  process,
  systems with different classes of  particles.  We also believe
  that the method followed here could be applied to the ASEP
  with open boundaries for which the Bethe equations have been 
  derived  recently \cite{essler,essler2}.
  For the open TASEP with all rates equal to one,
  it is known  from  the Matrix method  that the mean  stationary 
  current is  given by the ratio 
   of two consecutive Catalan numbers \cite{DEHP}: 
  can  this rather simple result be derived from  Bethe  Ansatz?

\subsection*{Acknowledgments} 
  We thank  Olivier Golinelli for many helpful  discussions and S. Mallick
 for a careful reading of the manuscript.

\appendix

\section{Derivation  of equation~(\ref{eq:solB2}) }
 \label{AppendixSOLQ2}

  We want to derive the formula~(\ref{eq:solB2}) for  $\tilde{B_2}(T)$
  which  is a particular solution of  equation~(\ref{eq:defB2modTN}). 
  We   substitute  the  formal expression~(\ref{def:tildeB2}) 
 of $\tilde{B_2}(T)$
 in  equation~(\ref{eq:defB2modTN}) and use the known
 explicit formulae for  $Q_1(T)$ and $R_1(T)$ (given in 
 equations~(\ref{solQ1}) and~(\ref{solR1}),  respectively).
 After identifying
 the terms of the same degree in $T$, the following linear system of
  equations is  obtained: 
\begin{eqnarray}
  \sum_{\substack{k+p =m \\0 \le k \le N-1 \\ 0 \le p \le L} } 
 \left(\begin{array}{c} L\\ p \end{array} \right)(x^p - x^{k}) c_k
  = 
\sum_{\substack{k+p+r =m +N \\ 0 \le k,p \le N-1 \\ 0 \le r \le L} } 
 \left(\begin{array}{c} L\\ r \end{array} \right)
    \left(\begin{array}{c} L\\ p \end{array} \right)
  \left(\begin{array}{c} L\\ k \end{array} \right)
\frac{x^p - x^r}{1 - x^{N-k}}
 - \left(\begin{array}{c} L\\ N \end{array} \right)
   \sum_{\substack{k+p =m \\ 0 \le k \le N-1 \\  0 \le p \le L} } 
  \left(\begin{array}{c} L\\ p \end{array} \right)
  \left(\begin{array}{c} L\\ k \end{array} \right)x^{k}
 \, ,
 \label{eqck}
 \end{eqnarray}
 where we have introduced
\begin{equation}
    c_k =   (-1)^{N+k+1} 
  \frac{\left(\begin{array}{c} L\\ N \end{array} \right)^2}{L^2}
   (1 - x^{N-k}) \,  q_k^{(2)} \, .
 \label{defck} 
\end{equation}
The system~(\ref{eqck}) is a triangular system of $N$ equations, parametrized
 by the integer $m$, with $0 \le m \le N-1\,.$
 In equation~(\ref{eqck}),  we have written explicitely the ranges
 for all the dummy variables.  However, some pieces of information 
 are  redundant:
 for example, we know that  $0 \le m \le N-1$;  therefore,  if $k+p =m$
  then both $k$ and $p$  must lie  between 0 and $N-1$
 (recall that a  binomial coefficient  with a negative entry  is  equal to 0).
 In the following, we shall  not write such  superfluous information.

 We now  start by transforming the r.h.s. of equation~(\ref{eqck}):
 we notice that the first 
 sum on the r.h.s. is formally antisymmetric with respect to the indices 
 $r$ and $p$. However, this sum does not  vanish identically  because
 the range of these two  variables is not the same. If the range of  $r$ 
  were from  0 to  $N-1$,  the total  sum would be equal to zero. 
 In other words, the terms   in the range 
  $0 \le r \le N-1$ do not contribute to the sum, only the terms with
 $N \le r \le  L$ contribute. This sum  is thus given by
\begin{eqnarray}
\sum_{r =N}^{L}
  \sum_{\substack{0 \le k,p \le N-1 \\k+p+r =m +N}} 
  \left(\begin{array}{c} L\\ r \end{array} \right)
    \left(\begin{array}{c} L\\ p \end{array} \right)
  \left(\begin{array}{c} L\\ k \end{array} \right)
\frac{x^p - x^r}{1 - x^{N-k}}   
 =  \sum_{r =0}^{L-N}\left(\begin{array}{c} L\\ N+r \end{array} \right)
 \sum_{\substack{k+p+r =m  \\ 0 \le k,p \le N-1 }}
   \left(\begin{array}{c} L\\ p \end{array} \right)
  \left(\begin{array}{c} L\\ k \end{array} \right)
\frac{x^p - x^{N+r}}{1 - x^{N-k}}    \nonumber \\
 =      \sum_{r =0}^{L-N} 
 \left(\begin{array}{c} L\\ N+r \end{array} \right)
   \sum_{k+p =m}
   \left(\begin{array}{c} L\\ p \end{array} \right)
  \left(\begin{array}{c} L\\ k -r \end{array} \right)
   \left( x^k + 
 \frac{x^p - x^k}{1 - x^{N + r -k}} \right)
  \label{somme2}
 \end{eqnarray}
 where, 
   we have first replaced the dummy variable $r$
 by $r-N$ and  then,  to derive  the last 
  equality,  we   use  the identity 
 $(x^p - x^{N+r})/(1 - x^{N-k}) =  x^{k+r} + (x^p - x^{k+r})/(1 - x^{N-k})$
 and  replace    $k$ by $k-r$.  
  Thus,  we rewrite the r.h.s.  of equation~(\ref{eqck}) as follows:
\begin{eqnarray}
     \sum_{r =0}^{L-N} 
 \left(\begin{array}{c} L\\ N+r \end{array} \right)
   \sum_{k+p =m}
   \left(\begin{array}{c} L\\ p \end{array} \right)
  \left(\begin{array}{c} L\\ k -r \end{array} \right)
   \left( x^k + 
 \frac{x^p - x^k}{1 - x^{N + r -k}} \right)
 - \left(\begin{array}{c} L\\ N \end{array} \right)
  \sum_{ k+p =m} 
  \left(\begin{array}{c} L\\ p \end{array} \right)
  \left(\begin{array}{c} L\\ k \end{array} \right)x^{k} \nonumber  \\
 =  \sum_{r =1}^{L-N} 
 \left(\begin{array}{c} L\\ N+r \end{array} \right)
   \sum_{ k+p =m}
   \left(\begin{array}{c} L\\ p \end{array} \right)
  \left(\begin{array}{c} L\\ k -r \end{array} \right) x^k
  +    \sum_{r =0}^{L-N} 
 \left(\begin{array}{c} L\\ N+r \end{array} \right)
   \sum_{k+p =m}
   \left(\begin{array}{c} L\\ p \end{array} \right)
  \left(\begin{array}{c} L\\ k -r \end{array} \right)
 \frac{x^p - x^k}{1 - x^{N + r -k}}  \, . 
\label{newrhs}
 \end{eqnarray}
  The first term   in the last
 equality  is now  rewritten  using  the following identity
\begin{eqnarray}
    \sum_{ k+p =m}
   \left(\begin{array}{c} L\\ p \end{array} \right)
  \left(\begin{array}{c} L\\ k -r \end{array} \right) x^k =
 \sum_{ k+p =m}
   \left(\begin{array}{c} L\\ p \end{array} \right)
  \left(\begin{array}{c} L\\ k -r \end{array} \right) 
  \frac{x^p - x^k}{1 - x^r} x^r \, .  
 \label{idfine}
\end{eqnarray}
 [This identity is readily proved after  multiplying both sides
 by $(1 - x^r)$, cancelling the common $x^{k+r}$  term and
 noticing that the remaining terms are identical up to a
 notation change.]

 Finally, 
  the initial  system~(\ref{eqck}) becomes:
\begin{eqnarray}
  \sum_{ k+p =m}
 \left(\begin{array}{c} L\\ p \end{array} \right)(x^p - x^{k}) c_k = 
 {\hskip 12 cm} \label{eqck2}   \\ 
    \sum_{r =1}^{L-N} 
 \left(\begin{array}{c} L\\ N+r \end{array} \right)
 \sum_{ k+p =m}
   \left(\begin{array}{c} L\\ p \end{array} \right)
  \left(\begin{array}{c} L\\ k -r \end{array} \right) 
  \frac{x^p - x^k}{1 - x^r} x^r
 +  \sum_{r =0}^{L-N} 
 \left(\begin{array}{c} L\\ N+r \end{array} \right)
   \sum_{k+p =m}
   \left(\begin{array}{c} L\\ p \end{array} \right)
  \left(\begin{array}{c} L\\ k -r \end{array} \right)
 \frac{x^p - x^k}{1 - x^{N + r -k}} 
\, .  \nonumber 
 \end{eqnarray}

 Clearly,  the solution of this equation is given by
\begin{eqnarray}
 c_k  = \sum_{r=1}^{N-1} 
 \frac{ x^{r}    \left(\begin{array}{c} L\\ N + r \end{array} \right)
  \left(\begin{array}{c} L\\ k- r \end{array} \right)  }
      { 1 - x^{r} } + 
    \sum_{r=0}^{N-1} 
  \frac{\left(\begin{array}{c} L\\ N + r \end{array} \right)
   \left(\begin{array}{c} L\\ k- r \end{array} \right) }
 { 1 - x^{N+r -k}}   
  \, .
\end{eqnarray}
  This ends the proof of the  formula~(\ref{eq:solB2}).


\section{Some useful steps in the calculation of $\Delta$}
 \label{AppendixUsteps}

\subsection{Binomial formulae}
 
 In the sequel, we shall use  repeatedly the following 
  elementary   binomial formulae:
 \begin{eqnarray}
 p  \left(\begin{array}{c} L\\ p \end{array} \right)  &=&
   L {\left(\begin{array}{c} L -1\\ p -1 \end{array} \right)}  \,  , 
 \label{Bin1}   \\
   (L-p){\left(\begin{array}{c} L\\ p \end{array} \right)}  &=& 
  L \left(\begin{array}{c} L -1 \\ p \end{array} \right)
  \label{Bin2} \, .  \\
    \sum_{p  = A}^B (-1)^{p}
      \left(\begin{array}{c} L\\ p \end{array} \right) 
 &=&  \sum_{p  = A}^B (-1)^{p} \left\{
    \left(\begin{array}{c} L-1\\ p \end{array} \right) 
 +  \left(\begin{array}{c} L-1\\ p -1 \end{array} \right) \right\}
  =  (-1)^{B} \left(\begin{array}{c} L-1\\ B \end{array} \right)
 + (-1)^{A} \left(\begin{array}{c} L-1\\ A-1 \end{array} \right) 
\label{Bin3}   \, , \\
 \sum_{r=0}^{N} r \left(\begin{array}{c} L\\ N +r  \end{array}\right) 
  \left(\begin{array}{c} L\\ N -r \end{array}\right)  &=& 
   \frac{L}{2} \sum_{r=0}^{N}
 \left\{   \left(\begin{array}{c} L-1\\ N +r-1  \end{array}\right) 
  \left(\begin{array}{c} L-1\\ N -r \end{array}\right) -
\left(\begin{array}{c} L-1\\ N +r  \end{array}\right) 
  \left(\begin{array}{c} L-1\\ N -r-1 \end{array}\right)
 \right\}   \nonumber \\    &=&  \frac{L}{2} 
     \left(\begin{array}{c} L-1\\ N -1 \end{array}\right)
     \left(\begin{array}{c} L-1\\ N \end{array}\right) 
 = \frac{N(L-N)}{2L}  
   \left(\begin{array}{c} L\\ N \end{array}\right)^2
  \label{Bin4}   \, . 
  \end{eqnarray}

\subsection{An expression for the diffusion constant}

  We  start   with equation~(\ref{primeQR1}) and expand the polynomials
 $Q$ and $R$ to the second order in $\gamma$.
 This allows us to derive the following expression
 for   $R_2'(1)$. We obtain
\begin{eqnarray} 
 Q_0(1)  R_2'(1) +  Q_1(1)  R_1'(1)
   + Q_2(1) R_0'(1) +  Q_0'(1) R_2(1) +  Q_1'(1) R_1(1)
 +  Q_2'(1) R_0(1) \nonumber \\
 =  -Lx^{N+1} (1 -   x)^{L-1}Q_2\left(\frac{1}{x}\right)
   + x^{N-1}(1 -   x)^L  Q_2'\left(\frac{1}{x}\right) \, .
\end{eqnarray}
 We know from equation~(\ref{eq:ordre0}) that $Q_0(1)=1$, $Q_0'(1)=N,$
 $R_0(1) = (1 -x)^L$.
 From  equation~(\ref{eq:auxil2}), we deduce  $R_1(1) = N(1 -x)^L$
   and $R_2(1) = N^2(1 -x)^L/2$.
  Finally, equations~(\ref{valR0p1}) and~(\ref{valR1p1})
 give  the values  of $R_0'(1)$  and $R_1'(1)$.
 We also use equation~(\ref{eq:auxilordre2}) to express
 $Q_2(1/x)$ in terms of  $Q_1(1)$  and  $Q_2(1)$.
 Substituting this information into the previous expression leads to
  (remark that terms proportional  to  $Q_2(1)$ cancel out):
 \begin{equation} 
 \frac{ R_2'(1)}{ (1 -x)^L}  =  -\frac{N^3}{2}
  - \frac{N^2 L x}{2(1-x)}  +  N \frac{L-N}{L-1} Q_1(1) - N Q_1'(1)
 -  Q_2'(1) +   x^{N-1}  Q_2'\left(\frac{1}{x}\right) \, .
\end{equation}
 Inserting this expression into the formula~(\ref{DeltaR2}) for 
 $\Delta$ gives
\begin{equation}
    \frac{\Delta}{2p(1-x)} = \frac{N^3}{2} -   N^2 \frac{L-N}{L-1}
 -  N \frac{L-N}{L-1} Q_1(1) +  N Q_1'(1) + B_2'(1) \, ,
\end{equation}
 where we have used the definition~(\ref{eq:defB2})
  of the polynomial $B_2(T)$.  With the help of equations~(\ref{eq:exprB2})
 (\ref{eq:solC}), and~(\ref{Bp1of1}) we get  
  \begin{equation}
  B_2'(1) = \tilde{B_2}'(1) +  B_1'(1)
  \left(\frac{\tilde{B_2}(1)}{N} + Q_1(1) + \frac{N}{2} \right)
 = \tilde{B_2}'(1)  -L \frac{N-1}{L-1} \tilde{B_2}(1)
    -LN \frac{N-1}{L-1} Q_1(1) - LN^2 \frac{N-1}{2(L-1)} \, . 
 \label{eq:exprdB2of1}
 \end{equation}
   Substituting this expression in equation~(\ref{DeltaApp}),
  we conclude that
\begin{equation} 
 \frac{\Delta}{2p(1-x)} = -N^2 \frac{L-N}{2(L-1)}-N^2 Q_1(1) + N Q_1'(1)
 +   \tilde{B_2}'(1)  -L\frac{N-1}{L-1}\tilde{B_2}(1) \, .
    \label{DeltaApp}
\end{equation}
 The values  of all the terms that appear in this equation
 are  known. We now evaluate each  of these terms separately.

\subsection{Calculation of some exact expressions}

   
  The value of $Q_1(1)$ is  easily obtained
 from the expression~(\ref{solQ1}) of $Q_1(T)$:
 \begin{eqnarray}
 Q_1(1) =  \frac{ (-1)^{N} L}
  {\left(\begin{array}{c} L\\ N \end{array} \right)} \sum_{r=0}^{N-1} 
  \frac{ (-1)^{r} \left(\begin{array}{c} L\\ r \end{array} \right)}
     { 1 - x^{N-r}} = \frac{L}
  {\left(\begin{array}{c} L\\ N \end{array} \right)} \sum_{r=1}^{N} 
   \frac{ (-1)^{r} \left(\begin{array}{c} L\\ N-r \end{array} \right)}
     { 1 - x^r}
 \,.  \label{Q1de1} 
\end{eqnarray}
 Similarly, we have 
\begin{eqnarray}
    Q_1'(1) =  \frac{L}
  {\left(\begin{array}{c} L\\ N \end{array} \right)} \sum_{r=1}^{N} 
 \frac{ (-1)^{r}  (N-r) \left(\begin{array}{c} L\\ N-r \end{array} \right)}
     { 1 - x^r}  \,,  \label{Q1primede1}
\end{eqnarray}

   To calculate $\tilde{B_2}(1)$,  we start from the formula
 for  $\tilde{B_2}(T)$ given in equations~(\ref{def:tildeB2})
  and~(\ref{eq:solB2}):
\begin{equation}
     \tilde{B_2}(1) =  
  \frac{ (-1)^{N+1} L^2}{\left(\begin{array}{c} L\\ N \end{array} \right)^2}
  \sum_{k=0}^{N-1}(-1)^{k}   \left\{
 \sum_{r=1}^{N-1} \,\,  
 \frac{ x^{r}  \left(\begin{array}{c} L\\ N + r \end{array} \right)
   \left(\begin{array}{c} L\\ k- r \end{array} \right) }
     { 1 - x^{r} }  +
    \sum_{r=0}^{N-1} 
  \frac{\left(\begin{array}{c} L\\ N + r \end{array} \right)
   \left(\begin{array}{c} L\\ k- r \end{array} \right) }
     { 1 - x^{N+r -k}}  
    \right\} \, .
 \label{B2de1}
\end{equation}
   Exchanging the double sum and using 
 equations~(\ref{Bin1}) and~(\ref{Bin3}), we rewrite
 the   first term   on the r.h.s  of this expression  as follows:  
\begin{equation}
 \sum_{r=1}^{N-1} \,\, 
  \frac{ x^{r}  \left(\begin{array}{c} L\\ N + r \end{array} \right)}
     { 1 - x^{r} } \sum_{k=0}^{N-1}(-1)^{k}
 \left(\begin{array}{c} L\\ k- r \end{array} \right)  =  
 \frac{ (-1)^{N-1}}{L} \sum_{r=1}^{N-1} \,\,  
     \frac{ x^{r} (N-r) \left(\begin{array}{c} L\\ N + r \end{array} \right)
   \left(\begin{array}{c} L\\ N -  r \end{array} \right)}  { 1 - x^{r} }  \, .
 \label{eq:transfo1}
\end{equation}
   In the  second term on the r.h.s. of~(\ref{B2de1}),
   we remark that  the  effective range of the variable  $r$ is 
  from  $0$ to $ k$ and we replace $r$
    by   $r' = k -r$. We then transform this term
   in  a  manner similar  to that described  in  equation~(\ref{eq:transfo1}).
    Finally, the expression~(\ref{B2de1}) simplifies to:
  \begin{equation}
     \tilde{B_2}(1)  =  
 \frac{ L}{\left(\begin{array}{c} L\\ N \end{array} \right)^2}
    \sum_{r=1}^{N}  
   \frac{  x^{r}(N-r) + N+r}  { 1 - x^{r} }  
   \left(\begin{array}{c} L\\ N + r \end{array} \right)
   \left(\begin{array}{c} L\\ N -  r \end{array} \right)
 -   \frac{ LN}{\left(\begin{array}{c} L\\ N \end{array} \right)}
    \sum_{r=1}^{N}  
  \frac{ (-1)^r \left(\begin{array}{c} L\\ N -  r \end{array} \right)} 
    { 1 - x^{r} }        \,  .
 \label{B2de1bis}
\end{equation}

   Using  similar steps, we find that  $ \tilde{B_2}'(1)$ 
 is given by
   \begin{eqnarray}
     \tilde{B_2}'(1) &=& 
  \frac{L}{(L-1)\left(\begin{array}{c} L\\ N \end{array} \right)^2}
     \sum_{r=1}^{N}  
   \frac{  x^{r}(N-r)(LN-r-L) + (N+r)(LN+r-L)}  { 1 - x^{r} }  
   \left(\begin{array}{c} L\\ N + r \end{array} \right)
   \left(\begin{array}{c} L\\ N -  r \end{array} \right) \nonumber \\
  &-&  \frac{ LN}{(L-1) \left(\begin{array}{c} L\\ N \end{array} \right)}
    \sum_{r=1}^{N}  
  \frac{ (-1)^r (LN-Lr+r-L)
  \left(\begin{array}{c} L\\ N -  r \end{array} \right)} 
    { 1 - x^{r} }    \, . 
 \label{B2primede1bis}
\end{eqnarray}

  To conclude our calculation, we must substitute equations~(\ref{Q1de1})
   (\ref{Q1primede1}),   (\ref{B2de1bis}), and~(\ref{B2primede1bis})
    into the formula~(\ref{DeltaApp}) for the diffusion constant.
      We find  that all the terms that contain only one binomial factor
     {\it i.e.}, terms proportional to 
 ${ (-1)^{r} \left(\begin{array}{c} L\\ N-r \end{array} \right)}/
     {(1 - x^r)}$   cancel out amongst themselves.
     After some elementary simplications, we are left with
\begin{equation} 
 \frac{\Delta}{p(1-x)} = 
 \frac{2L}{(L-1)\left(\begin{array}{c} L\\ N \end{array}\right)^2 }
   \sum_{r=1}^{N} r^2 \frac{ 1 + x^r}{1 - x^r} 
 \left(\begin{array}{c} L\\ N +r  \end{array}\right) 
  \left(\begin{array}{c} L\\ N -r \end{array}\right)  + 
\frac{2 LN}{(L-1) \left(\begin{array}{c} L\\ N \end{array} \right)^2}
   \sum_{r=0}^{N} r \left(\begin{array}{c} L\\ N +r  \end{array}\right) 
  \left(\begin{array}{c} L\\ N -r \end{array}\right)
 - N^2 \frac{L-N}{(L-1)}
\end{equation}
            The   last  two terms cancel with each other 
 according to the identity~(\ref{Bin4}).
 This ends the proof of equation~(\ref{formuleCstedeDiff}).

\section{Functional Bethe Ansatz for TASEP}

 We  consider here  the special  case of the TASEP 
 (which corresponds to $p=1$ and   $q =x=  0$). We explain how
 to retrieve from   the Q-R equation (\ref{FBA})
 the parametric representation of $E_{{\rm max}}(\gamma)$ that was  obtained
 in   \cite{DLebowitz} by using contour integrals.
For   $x=  0$,  the functional equation (\ref{FBA}) reduces to
\begin{equation}
Q(T)R(T)=T^{N}+ (-1)^{N-1} B (1-T)^{L} \,\,\, \hbox{ with  } \,\,\,
B=(-1)^{N-1}e^{L\gamma}Q(0)  \, . 
 \label{FBATASEP}
\end{equation}
From equation~(\ref{eq:ordre0}), we find  that 
 the zeroth order polynomials for the TASEP are simply given by
\begin{eqnarray}
 Q_0(T) =  T^N \, \,\,\,\,\,\,  \hbox{ and }
 {\hskip 1cm}   R_0(T) =  1   \, . 
  \label{eq:ordre0TASEP}
\end{eqnarray}
 The perturbative expansions~(\ref{eq:devQ}) and~(\ref{eq:devR}) can
  be rewritten as  
\begin{equation}
      Q(T) =   T^N + \gamma {\mathcal Q}(T) \,\,\, \hbox{ and  } \,\,\,
         R(T)  = 1 + \gamma {\mathcal R}(T) \, ,
   \label{devTASEP}
\end{equation}
where ${\mathcal Q}(T)$ is  a polynomial  of degree  $N-1$ 
 and  ${\mathcal R}(T)$ is  of degree  $L-N$ 
 (the coefficients of these two polynomials  are functions of  $\gamma$).
 We note, in particular, that 
 $Q(0)$ is of  order  $\gamma$ and  thus  $B$ defined in 
  equation~(\ref{FBATASEP}) is also of order  $\gamma$ 
 and is   a  small parameter.
Dividing both sides of equation~(\ref{FBATASEP})
 by $T^{N}$ and taking the logarithm, we obtain 
\begin{equation}
\log\left(\frac{Q(T)}{T^{N}}\right)+\log R(T)
=\log\left(1+ (-1)^{N-1} B \frac{(1-T)^{L}}{T^{N}}\right)
=\sum_{k=1}^{\infty}\frac{(-1)^{Nk-1} B^{k} }{k}
\frac{(1-T)^{kL}}{T^{kN}}  \, , 
\label{TASEPlogQlogR}
\end{equation}
where we  have developped the logarithm in powers of $B$. We remark that
   the r.h.s.  of this equation
 is a series that contains   both positive and negative powers of  $T$.
  But,  equation~(\ref{devTASEP})  implies that 
  ${Q(T)}/{T^{N}} = 1 + \gamma {\mathcal Q}(T)/T^N$, {\it i.e.}, 
  ${Q(T)}/{T^{N}}$   is a polynomial in the variable $1/T$  of degree $N$.
 Therefore,  the expansion of  $\log\left({Q(T)}/{T^{N}}\right)$
 w.r.t. $\gamma$ (or $B$) can only generate negative powers of $T$.
 Similarly,   from equation~(\ref{devTASEP})
 we have  $\log R(T) = \log( 1 + \gamma {\mathcal R}(T))$
 and the expansion of this term can  generate only  positive 
 powers of $T$. Therefore, the identification between
 the l.h.s. and the  r.h.s.  of  equation~(\ref{TASEPlogQlogR}) 
  is unique and we have 
\begin{eqnarray} 
\log\left(\frac{Q(T)}{T^{N}}\right)  &=&
 \sum_{k=1}^{\infty}\frac{(-1)^{Nk-1} B^k}{k}
\sum_{j=0}^{kN-1}(-1)^{j}
\left(\begin{array}{c}kL\\j\end{array}\right)T^{j-kN}
\label{TASEPlogQ}           \\
\log R(T)  &=& \sum_{k=1}^{\infty}\frac{(-1)^{Nk-1} B^k}{k}
\sum_{j=kN}^{kL}(-1)^{j}
\left(\begin{array}{c}kL\\j\end{array}\right)T^{j-kN}
\label{TASEPlogR}
\end{eqnarray}

 For the TASEP,  equations~(\ref{eq:vp3}) and ~(\ref{eq:auxil2})
 reduce to
 \begin{equation}
       E_{{\rm max}}(\gamma) =   - \frac{ R'(1)}{ R(1)}  
   = - \frac{d}{dT}\log R(T) \Big|_{T=1}   \,\,\, \hbox{ and }
  \,\,\,    \gamma  = \frac{1}{N} \log R(1) \, . 
 \end{equation}
   From equation~(\ref{TASEPlogR}), we obtain 
 (with the help of equations
 (\ref{Bin1}) and (\ref{Bin3}) to calculate  the sums over $j$):
\begin{eqnarray}
 E_{{\rm max}}(\gamma) &=&
 -N\sum_{k=1}^{\infty}B^{k}\frac{(kL-2)!}{(kN)!(kL-kN-1)!}  \, ,  \\
 \gamma  &=& -\sum_{k=1}^{\infty}B^{k}\frac{(kL-1)!}{(kN)!(kL-kN)!} \, .
\end{eqnarray}
These  two equations are precisely those derived in \cite{DLebowitz}. They 
 provide  a parametric formula for $E_{{\rm max}}(\gamma)$ that
 allows  to calculate the large deviation function of the current
  and   its  cumulants to any required order.

\end{document}